\documentclass[rnaas]{aastex63}

\usepackage{amsmath}
\usepackage{afterpage}
\usepackage{scalerel}
\usepackage{natbib}
\bibpunct[; ]{(}{)}{;}{a}{}{,}
\usepackage{multirow}
\usepackage{graphics}
\usepackage{threeparttable}
\usepackage[varg]{txfonts}
\usepackage{xcolor}
\usepackage{bm}
\hypersetup{colorlinks,linkcolor={blue},citecolor={blue},urlcolor={blue}}

\begin{document}
\title{Photometric Redshifts in the W-CDF-S and ELAIS-S1 Fields Based on Forced Photometry from 0.36 -- 4.5 Microns}

\author[0000-0002-4436-6923]{Fan Zou}
\affiliation{Department of Astronomy and Astrophysics, 525 Davey Lab, The Pennsylvania State University, University Park, PA 16802, USA}
\affiliation{Institute for Gravitation and the Cosmos, The Pennsylvania State University, University Park, PA 16802, USA}

\author[0000-0001-8835-7722]{Guang Yang}
\affiliation{Department of Physics and Astronomy, Texas A$\&$M University, College Station, TX, 77843-4242 USA}
\affiliation{George P. and Cynthia Woods Mitchell Institute for Fundamental Physics and Astronomy, Texas A$\&$M University, College Station, TX, 77843-4242 USA}

\author[0000-0002-0167-2453]{W. N. Brandt}
\affiliation{Department of Astronomy and Astrophysics, 525 Davey Lab, The Pennsylvania State University, University Park, PA 16802, USA}
\affiliation{Institute for Gravitation and the Cosmos, The Pennsylvania State University, University Park, PA 16802, USA}
\affiliation{Department of Physics, 104 Davey Laboratory, The Pennsylvania State University, University Park, PA 16802, USA}

\author[0000-0002-8577-2717]{Qingling Ni}
\affiliation{Department of Astronomy and Astrophysics, 525 Davey Lab, The Pennsylvania State University, University Park, PA 16802, USA}
\affiliation{Institute for Gravitation and the Cosmos, The Pennsylvania State University, University Park, PA 16802, USA}

\author[0000-0002-8686-8737]{Franz E. Bauer}
\affiliation{Instituto de Astrof\'isica, Facultad de F\'isica, Pontificia Universidad Cat\'olica de Chile Av. Vicu\~na Mackenna 4860, 782-0436 Macul, Santiago, Chile}
\affiliation{National Radio Astronomy Observatory, Pete V. Domenici Array Science Center, P.O. Box O, Socorro, NM 87801, USA}
\affiliation{Space Science Institute, 4750 Walnut Street, Suite 205, Boulder, CO 80301, USA}

\author[0000-0002-2553-096X]{Giovanni Covone}
\affiliation{INAF - Osservatorio Astronomico di Capodimonte, Salita Moiariello 16, I-80131, Napoli, Italy}
\affiliation{Dipartimento di Fisica, Universit\`a di Napoli ``Federico II'', via Cinthia 9, 80126 Napoli, Italy}
\affiliation{INFN - Sezione di Napoli, via Cinthia 9, 80126 Napoli, Italy}

\author[0000-0002-3032-1783]{Mark Lacy}
\affiliation{National Radio Astronomy Observatory, 520 Edgemont Road, Charlottesville, VA 22903, USA}

\author[0000-0003-0911-8884]{Nicola R. Napolitano}
\affiliation{INAF - Osservatorio Astronomico di Capodimonte, Salita Moiariello 16, I-80131, Napoli, Italy}

\author[0000-0003-1991-370X]{Kristina Nyland}
\affiliation{National Research Council, resident at the U.S. Naval Research Laboratory, 4555 Overlook Ave. SW, Washington, DC 20375, USA}

\author[0000-0003-4210-7693]{Maurizio Paolillo}
\affiliation{Dipartimento di Fisica, Universit\`a di Napoli ``Federico II'', via Cinthia 9, 80126 Napoli, Italy}
\affiliation{INFN - Sezione di Napoli, via Cinthia 9, 80126 Napoli, Italy}
\affiliation{INAF - Osservatorio Astronomico di Capodimonte, Salita Moiariello 16, I-80131, Napoli, Italy}

\author[0000-0002-3585-866X]{Mario Radovich}
\affiliation{INAF - Osservatorio Astronomico di Padova, vicolo Osservatorio, 5 I-35122 Padova, Italy}

\author[0000-0002-6427-7039]{Marilena Spavone}
\affiliation{INAF - Osservatorio Astronomico di Capodimonte, Salita Moiariello 16, I-80131, Napoli, Italy}

\author[0000-0002-6748-0577]{Mattia Vaccari}
\affiliation{Inter-university Institute for Data Intensive Astronomy, Department of Physics and Astronomy, University of the Western Cape, Robert Sobukwe Road, 7535 Bellville, Cape Town, South Africa}
\affiliation{INAF - Istituto di Radioastronomia, via Gobetti 101, 40129 Bologna, Italy}

\email{E-mail: fuz64@psu.edu, gyang206265@gmail.com}

\begin{abstract}
The \mbox{W-CDF-S} and \mbox{ELAIS-S1} fields will be two of the LSST Deep Drilling fields, but the availability of spectroscopic redshifts within these two fields is still limited on $\mathrm{deg^2}$ scales. To prepare for future science, we use \texttt{EAZY} to estimate photometric redshifts (photo-\textit{z}s) in these two fields based on forced-photometry catalogs. Our photo-\textit{z} catalog consists of $\sim0.8$ million sources covering $4.9~\mathrm{deg^2}$ in \mbox{W-CDF-S} and $\sim0.8$ million sources covering $3.4~\mathrm{deg^2}$ in \mbox{ELAIS-S1}, among which there are $\sim0.6$ (\mbox{W-CDF-S}) and $\sim0.4$ (\mbox{ELAIS-S1}) million sources having signal-to-noise-ratio (SNR) $>5$ detections in more than 5 bands. By comparing photo-\textit{z}s and available spectroscopic redshifts, we demonstrate the general reliability of our photo-\textit{z} measurements. Our photo-\textit{z} catalog is publicly available at\dataset[10.5281/zenodo.4603178]{\doi{10.5281/zenodo.4603178}}.
\end{abstract}

\section*{}
The Wide Chandra Deep Field-South (\mbox{W-CDF-S}) and European Large-Area ISO Survey-S1 (\mbox{ELAIS-S1}) fields are parts of the XMM-SERVS (\citealt{Chen18}; Ni et al., in preparation)\footnote{\url{http://personal.psu.edu/wnb3/xmmservs/xmmservs.html}} and MIGHTEE \citep{Jarvis16} surveys, and will further be two of the Deep Drilling Fields (e.g., \citealt{Brandt18}) of the Vera C. Rubin Observatory Legacy Survey of Space and Time (LSST). Redshifts are a crucial ingredient for most extragalactic studies, but spectroscopic redshifts (spec-\textit{z}s) are only available for a small fraction of VIDEO-selected sources in these two fields --- 3.8\% for \mbox{W-CDF-S} and 1.3\% for \mbox{ELAIS-S1}, and the spectroscopic coverage reaches roughly down to $i$-band magnitude $\sim23$ for \mbox{W-CDF-S} and $\sim22$ for \mbox{ELAIS-S1}. To lay a necessary foundation for future science, we measure photometric redshifts (photo-\textit{z}s) for 1.6 million sources in these two fields utilizing forced-photometry catalogs in Nyland et al. (in preparation; \mbox{W-CDF-S}) and \citet[\mbox{ELAIS-S1}]{Zou21} with \texttt{EAZY} \citep{Brammer08}. These catalogs were generated with \textit{The Tractor} \citep{Lang16} and provide consistent, deblended photometry across the available bands and are thus expected to improve the reliability of photo-\textit{z} measurements (e.g., \citealt{Nyland17}; \citealt{Chen18}).\par
Our photo-\textit{z} derivation methodology is the same as those in \citet{Yang14} and \citet{Chen18}, and interested readers can refer to these two articles for a detailed description. Briefly, \texttt{EAZY} fits the spectral energy distributions (SEDs) with a linear combination of several galaxy templates at a series of redshift grid points to estimate the redshifts. We adopt the default v1.3 set, which includes 9 templates, and default parameter settings in \texttt{EAZY}. All the templates are for pure-galaxy SEDs and do not include any AGN or stellar templates. We iteratively adjust the photometric zero points so that the median value of the differences between the fitted and observed magnitudes is zero. Unlike \citet{Brammer08}, we do not apply a magnitude prior, because this configuration tends to underestimate redshifts \citep{Yang14}. \texttt{EAZY} also returns a parameter, $Q_z$, used to indicate photo-\textit{z} quality (see Eq.~8 of \citealt{Brammer08}). A nominal criterion for ``high-quality'' photo-\textit{z}s is $Q_z<1$, and we adopt this hereafter. For sources with $\mathrm{SNR>5}$ detections in more than 5 bands,\footnote{Note that photo-\textit{z}s are still estimated for all sources regardless of the SNRs.} the overall completeness of high-quality photo-\textit{z}s is 69\% (0.39 out of 0.57 million sources) and 68\% (0.28 out of 0.42 million sources) in \mbox{W-CDF-S} and \mbox{ELAIS-S1}, respectively.\par
\begin{figure*}
\centering
\resizebox{0.85\hsize}{!}{\includegraphics{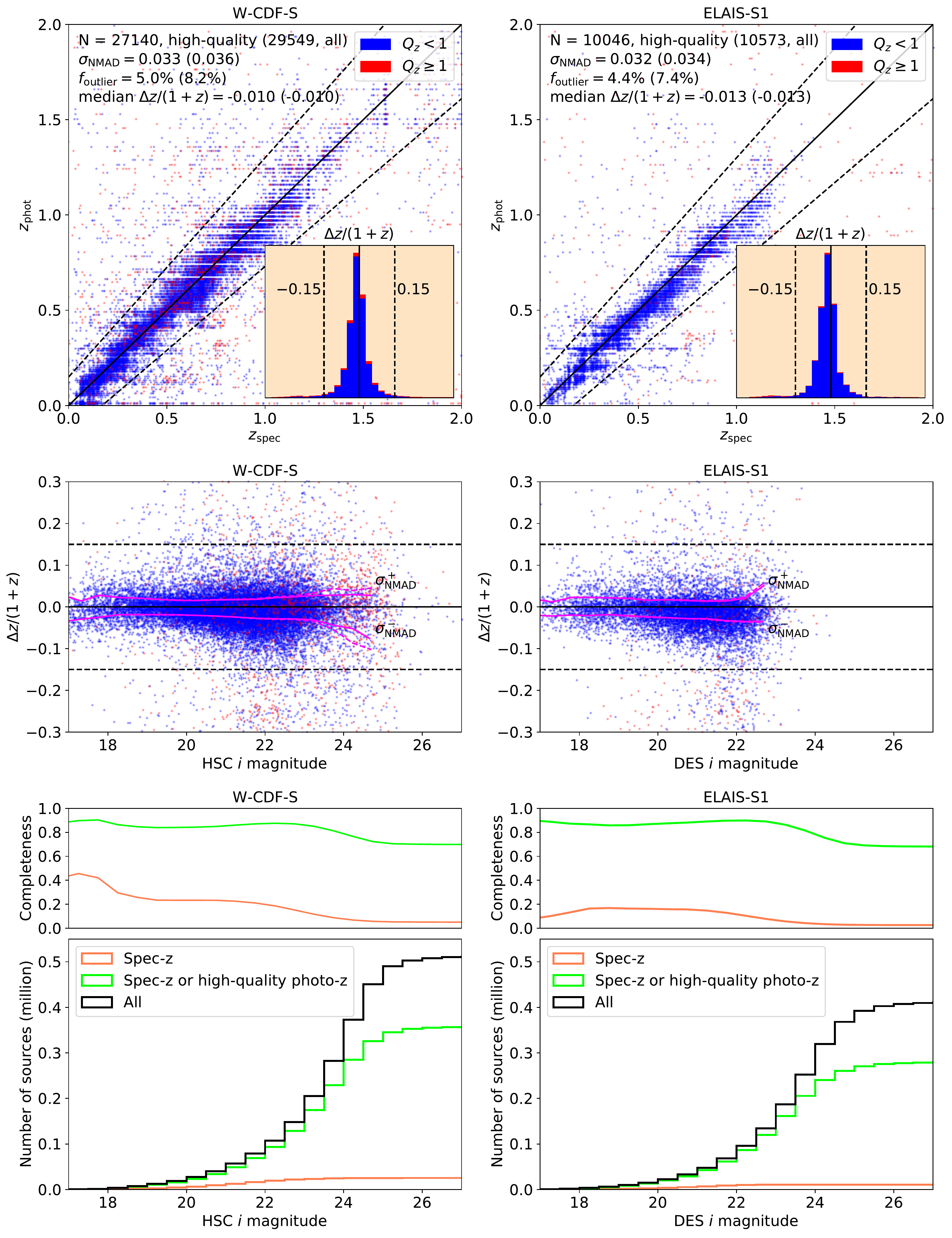}}
\caption{\textit{Top:} Comparisons between photo-\textit{z} and spec-\textit{z} values in \mbox{W-CDF-S} (\textit{left}) and \mbox{ELAIS-S1} (\textit{right}), where the photo-\textit{z} is the \textit{z}-grid value that minimizes the fitting $\chi^2$, and thus the possible photo-\textit{z} values are mildly ``quantized''. The blue and red colors represent sources with $Q_z<1$ and $Q_z\ge1$, respectively. The black solid lines mark $z_\mathrm{phot}=z_\mathrm{spec}$ relations, and the black dashed lines indicate $\Delta z/(1+z)=\pm0.15$, beyond which outliers are defined. The inset histograms show the distributions of $\Delta z/(1+z)$. Presented in the upper-left corners of the two panels are some diagnostic statistics, where $N$ is the number of sources. The values outside and inside the parentheses are for high-quality (i.e., $Q_z<1$) photo-\textit{z}s and all photo-\textit{z}s, respectively. \textit{Middle:} $\Delta z/(1+z)$ versus \textit{i}-band magnitudes. The fuchsia $\sigma_\mathrm{NMAD}^+$ and $\sigma_\mathrm{NMAD}^-$ curves are $\sigma_\mathrm{NMAD}$ and $-\sigma_\mathrm{NMAD}$ computed for sources with $\Delta z>0$ and $\Delta z\le0$, respectively. The fuchsia solid and dashed lines represent sources with $Q_z<1$ and all the sources, respectively. \textit{Bottom:} The cumulative distribution of the \textit{i}-band magnitudes. The coral and lime curves in the top sub-panels show the cumulative fractions of sources with spec-\textit{z}s and those with either high-quality photo-\textit{z}s or spec-\textit{z}s, respectively, among all the sources that have $\mathrm{SNR>5}$ detections in more than 5 bands as a function of \textit{i}-band magnitude.}
\label{Fig_photozmasterfig}
\end{figure*}

To evaluate the reliability of our photo-\textit{z}s, we compile spec-\textit{z}s for extragalactic sources, as summarized in Ni et al. (in preparation), who will release \textit{XMM-Newton} point-source catalogs in these fields. For the 29549 (\mbox{W-CDF-S}) and 10573 (\mbox{ELAIS-S1}) extragalactic sources that have $\mathrm{SNR>5}$ detections in more than 5 bands, we present a comparison between their spec-\textit{z}s and photo-\textit{z}s in Fig.~\ref{Fig_photozmasterfig}, in which we define $\Delta z=z_\mathrm{phot}-z_\mathrm{spec}$, $\sigma_\mathrm{NMAD}$ as the normalized median absolute deviation, and ``outliers'' as objects with $|\Delta z|/(1+z_\mathrm{spec})>0.15$. There are several factors that may cause outliers (see, e.g., Section~4.2 of \citealt{Brammer08}), though it is hard to know which is dominant. For 27140 sources with both spec-\textit{z}s and $Q_z<1$ photo-\textit{z}s in \mbox{W-CDF-S}, $\sigma_\mathrm{NMAD}=0.033$, outlier fraction ($f_\mathrm{outlier}$) $=5.0\%$, and median $\Delta z/(1+z_\mathrm{spec})=-0.010$. For \mbox{ELAIS-S1}, there are 10046 such sources, which have $\sigma_\mathrm{NMAD}=0.032$, $f_\mathrm{outlier}=4.4\%$, and median $\Delta z/(1+z_\mathrm{spec})=-0.013$. These values are better than those in \citet{Pforr19}; they derived photo-\textit{z}s for \textit{Spitzer}-detected sources based on unforced photometry in 12 bands in these two fields and obtained $\sigma_\mathrm{NMAD}>0.05$ and $f_\mathrm{outlier}>10\%$ in both fields. Additionally, our photo-\textit{z} results largely do not suffer from blending issues in DeepDrill because the adopted forced-photometry catalogs have already deblended the sources, and thus the $\sigma_\mathrm{NMAD}$ and $f_\mathrm{outlier}$ values are similar between blended and non-blended sources after matching their magnitudes. We also note that the median $\Delta z/(1+z_\mathrm{spec})$ offsets are much smaller than the typical uncertainties of the photo-\textit{z}s. Fig.~\ref{Fig_photozmasterfig} further shows that the overall photo-\textit{z} reliability decreases with increasing magnitude because fainter sources have larger photometric uncertainties and fewer photometric points. Since the depth of the spectroscopic coverage in \mbox{W-CDF-S} is deeper than that in \mbox{ELAIS-S1}, the exact values of $\sigma_\mathrm{NMAD}$ and $f_\mathrm{outlier}$ are not directly comparable between these two fields.\par

Our photo-\textit{z} catalog is available at\dataset[10.5281/zenodo.4603178]{\doi{10.5281/zenodo.4603178}} and provides spec-\textit{z}s (if available), photo-\textit{z}s, the 68\% lower and upper limits of photo-\textit{z}s, $Q_z$, numbers of bands with $\mathrm{SNR>5}$ detections, and the bands with the shortest and longest wavelengths among those with $\mathrm{SNR>5}$ detections. Our released data also include some supplementary notes for this catalog, in which we present more details about the adopted photometric data, estimate a nominal depth of high-quality photo-\textit{z}s to be $i$-band magnitude $\approx24$, give additional notes on Fig.~\ref{Fig_photozmasterfig}, and discuss the photo-\textit{z} uncertainties and caveats for AGNs and stars. Readers are encouraged to also refer to these supplementary notes for a better understanding of our data. Especially, since our templates lack AGN contributions, our photo-\textit{z}s may be less reliable for AGN-dominated sources, and these sources are often broad-line (BL) AGNs. Ni et al. (in preparation) select BL AGNs in these two fields and use templates and codes designed for fitting these sources to measure their photo-\textit{z}s. Users focusing on BL AGNs should use the photo-\textit{z} measurements in Ni et al. (in preparation) instead.

\bigbreak
\textit{Acknowledgements.} We acknowledge support from NASA grant 80NSSC19K0961. The National Radio Astronomy Observatory is a facility of the National Science Foundation operated under cooperative agreement by Associated Universities, Inc. Basic research in radio astronomy at the U.S. Naval Research Laboratory is supported by 6.1 Base Funding. FEB acknowledges support from ANID-Chile Basal AFB-170002, FONDECYT Regular 1200495 and 1190818, and Millennium Science Initiative -- ICN12\_009. MV acknowledges support from the Italian Ministry of Foreign Affairs and International Cooperation (MAECI Grant Number ZA18GR02) and the South African Department of Science and Technology's National Research Foundation (DST-NRF Grant Number 113121) as part of the ISARP RADIOSKY2020 Joint Research Scheme.

\newpage
\bibliography{citations}

\appendix

\section{The data}
\label{sec: the_data}
The utilized photometric bands for \mbox{W-CDF-S} are VOICE $ugri$ \citep{Vaccari16}, HSC $griz$ \citep{Ni19}, VIDEO $ZYJHK_s$ \citep{Jarvis13}, and DeepDrill 3.6 and $4.5~\mathrm{\mu m}$ \citep{Lacy21}; the bands for \mbox{ELAIS-S1} are VOICE $u$, DES DR2 $grizY$ \citep{Abbott21}, ESIS $BVR$ \citep{Berta06, Vaccari16}, VIDEO $ZYJHK_s$, and DeepDrill 3.6 and $4.5~\mathrm{\mu m}$. The photometric data are corrected for Galactic extinction following the method in Section~5.5 of \citet{Yang14}; see also \url{https://irsa.ipac.caltech.edu/applications/DUST/}. The compilation of spec-\textit{z}s is described in Ni et al. (in preparation), and interested readers can refer to their article for more details.\par
Our cataloged sources are required to be detected in VIDEO. However, the VIDEO survey in the $4.9~\mathrm{deg^2}$ W-CDF-S field is not uniform, and some regions are shallower than ELAIS-S1 or even are not covered in some bands; e.g., the $Z$ band only covers $1.8~\mathrm{deg^2}$ of the entire W-CDF-S field. Therefore, W-CDF-S has a slightly smaller surface number density on $\mathrm{deg^2}$ scales than ELAIS-S1.\par

\section{The depth of high-quality photo-\textit{z}s}
To assess quantitatively how the fraction of high-quality photo-\textit{z}s drops as a function of $i$-band magnitude ($i_\mathrm{mag}$), we fit our data with the following formulae:
\begin{align}
p(i_\mathrm{mag})&=\frac{a_1}{1+\exp{[a_2(i_\mathrm{mag}-a_3)]}}+a_4,\\
\delta_{Q_z<1}&\sim Bernoulli(p),
\end{align}
where ($a_1, a_2, a_3, a_4$) are parameters to be fitted, $p(i_\mathrm{mag})$ is the probability that a source with $i_\mathrm{mag}$ has $Q_z<1$, and $\delta_{Q_z<1}$ is defined as 1 if $Q_z<1$ and 0 if $Q_z\ge1$. $a_3$ can be regarded as a nominal magnitude, below which $p(i_\mathrm{mag})$ drops considerably (if $a_2>0$). By maximizing the corresponding likelihoods, we obtain
\begin{align}
p(i_\mathrm{mag})&=\frac{0.37}{1+\exp{[2.56(i_\mathrm{mag}-23.9)]}}+0.50~\textrm{for \mbox{W-CDF-S} with SNR cuts},\\
p(i_\mathrm{mag})&=\frac{0.63}{1+\exp{[1.87(i_\mathrm{mag}-24.3)]}}+0.23~\textrm{for \mbox{W-CDF-S} without SNR cuts},\\
p(i_\mathrm{mag})&=\frac{0.48}{1+\exp{[2.91(i_\mathrm{mag}-23.8)]}}+0.42~\textrm{for \mbox{ELAIS-S1} with SNR cuts},\\
p(i_\mathrm{mag})&=\frac{0.79}{1+\exp{[2.38(i_\mathrm{mag}-24.1)]}}+0.10~\textrm{for \mbox{ELAIS-S1} without SNR cuts},
\end{align}
where ``SNR cuts'' means that we only select sources having $\mathrm{SNR>5}$ detections in more than 5 bands.\par
Fig.~\ref{Fig_prob_highQ} justifies that these functions can generally match well the observed fraction of high-quality photo-\textit{z}s as a function of $i_\mathrm{mag}$. The figure also shows that applying the SNR cut can greatly improve the overall photo-\textit{z} quality in the faint-magnitude regime. The best-fit $a_3$ values are generally around 24, and thus we conclude that the nominal $i$-band ``depths'' of our high-quality photo-\textit{z}s are $\sim24$ for both \mbox{W-CDF-S} and \mbox{ELAIS-S1}. The $p(i_\mathrm{mag})$ values of \mbox{ELAIS-S1} can generally match those of \mbox{W-CDF-S} at $i_\mathrm{mag}\lesssim24$. At fainter magnitudes, the \mbox{ELAIS-S1} $p(i_\mathrm{mag})$ drops to a lower level than the \mbox{W-CDF-S} $p(i_\mathrm{mag})$ because the $griz$ data in \mbox{ELAIS-S1} are shallower.

\renewcommand\thefigure{S1}
\begin{figure*}[t]
\centering
\resizebox{\hsize}{!}{\includegraphics{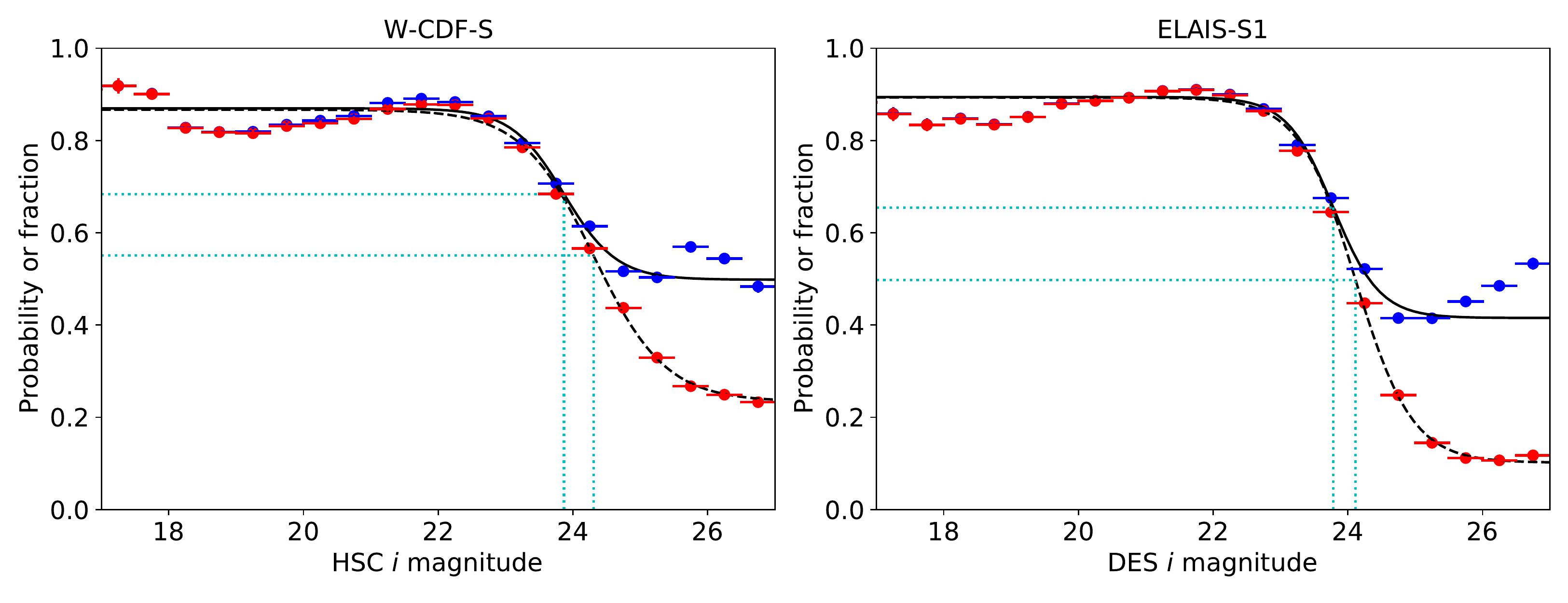}}
\caption{The points with error bars are the fractions of high-quality photo-\textit{z}s in magnitude bins with widths = 0.5 mag for \mbox{W-CDF-S} (\textit{left}) and \mbox{ELAIS-S1} (\textit{right}). The blue and red points represent cases with and without SNR cuts, respectively. The black curves are the best-fit probabilities, $p(i_\mathrm{mag})$, of obtaining a high-quality photo-\textit{z} for a source as a function of its $i$-band magnitude. The solid and dashed curves are for cases with and without SNR cuts, respectively. The cyan dotted lines mark the nominal depths (i.e., $a_3$) of our high-quality photo-\textit{z}s as well as the corresponding $p(i_\mathrm{mag})$ values. The points and curves are generally consistent.}
\label{Fig_prob_highQ}
\end{figure*}

\section{Additional Notes on Figure~1}
There are a few sources that have non-zero spec-\textit{z}s but have near-zero photo-\textit{z}s. This phenomenon may arise because the \texttt{EAZY} code cannot find an obvious Balmer break at 3646~\AA\ for these sources when no bands cover the break, and thus it assigns the photo-\textit{z}s to be $\sim0$.

\section{Notes on the photo-\textit{z} uncertainties}
The cataloged photo-\textit{z} is the \textit{z}-grid value that minimizes the fitting $\chi^2$ output by \texttt{EAZY}, and the 68\% limits are the redshift values of the photo-\textit{z} quantile function from the fitting at 16\% and 84\% (see Eq.~6 of \citealt{Brammer08}). The definition of the best-fit photo-\textit{z} does not require it to be necessarily within the photo-\textit{z} uncertainty interval; thus, there are a small fraction of sources with photo-\textit{z}s smaller/larger than their 68\% lower/upper limits, and the limits of these sources should not be used \citep{Yang14}. Furthermore, we found that 78\%/76\% of spec-\textit{z}s reside within the 68\% photo-\textit{z} intervals in \mbox{W-CDF-S}/\mbox{ELAIS-S1}. This may indicate that the photo-\textit{z} uncertainties are slightly overestimated (i.e., 78\%/76\% roughly correspond to 1.2 $\sigma$ for a normal distribution).

\section{Notes on AGNs and stars}
As mentioned in the main text, we only use galaxy templates to fit all the sources, and thus the results may be less reliable for AGNs and stars. For AGNs, however, only those that materially affect the observed optical-to-near-infrared (NIR) SEDs are expected to have this issue, and such sources are usually BL AGNs. Other sources with at most moderate AGN contributions to the optical-to-NIR SEDs (e.g., obscured AGNs) generally still have reliable photo-\textit{z}s. For example, Ni et al. (in preparation) show that $f_\mathrm{outlier}$ in \mbox{W-CDF-S}/\mbox{ELAIS-S1} are 6.8\%/4.4\% for non-BL X-ray AGNs with high-quality photo-\textit{z}s. Therefore, for most work involving AGNs in \mbox{W-CDF-S} and \mbox{ELAIS-S1}, to obtain optimal photo-\textit{z}s, it is usually sufficient to divide their sample to BL AGNs and other sources and adopt the photo-\textit{z}s in Ni et al. (in preparation) for the former and our photo-\textit{z}s for the latter. We emphasize that the term ``BL AGNs'' here does not only refer to those AGNs with spectroscopically detected broad lines, but is generalized to mean all the sources with significant characteristics that are similar to spectroscopic BL AGNs, such as those sources with AGN-dominated SEDs.\par
In addition, BL AGNs are not expected to be dominant among the outliers in Fig.~\ref{Fig_photozmasterfig} because BL AGNs have a much lower surface number density ($\lesssim300~\mathrm{deg^{-2}}$; Ni et al., in preparation) than galaxies. Only 5\%/16\% of the outliers in W-CDF-S/ELAIS-S1 are identified as BL AGN candidates (Ni et al., in preparation), and these values further drop to 3\%/5\% when requiring $Q_z<1$.\par
Detailed selections of BL AGNs and stars are beyond the scope of this work. Thus, we leave the selections to the users of this catalog who can apply their own criteria, and they can also adopt the selections in Ni et al. (in preparation) for both BL AGNs and stars.\par

\end{document}